\documentclass[12pt,preprint]{aastex}
\usepackage{amsmath}
\newcommand{\bea}{\begin{eqnarray} }
\newcommand{\eea}{\end{eqnarray}}
\def\mbf#1{\mbox{\boldmath ${#1}$}}
\newcommand{\mpc}{$M_\odot$ pc$^{-2}\,$}

\begin{document}

\title{Gravity-driven Turbulence in Galactic Disks}

\author{Keiichi WADA$^{1,2}$, Gerhardt MEURER$^{3}$ AND Colin A. NORMAN$^{3,4}$}
\altaffiltext{1}{National Astronomical Observatory of Japan, Mitaka, Tokyo
181-8588, Japan\\
E-mail: wada.keiichi@nao.ac.jp}
\altaffiltext{2}{CASA, University of Colorado, 389UCB, Boulder, CO 80309-389}
\altaffiltext{3}{Johns Hopkins University, Baltimore, MD 21218}
\altaffiltext{4}{Space Telescope Science Institute, Baltimore, MD 21218}

\begin{abstract}
High-resolution, 2-D hydrodynamical simulations with a large dynamic
range are performed to study the turbulent nature of the interstellar
medium (ISM) in galactic disks. 
The simulations are global, where the self-gravity of the
ISM, realistic radiative cooling, and galactic rotation are taken into
account.
In the analysis undertaken here, feedback processes from stellar
energy source are omitted. 
 We find that the velocity field of the disk in a non-linear
phase shows a steady power-law energy spectrum over
three-orders of magnitude in wave number.
This implies that the random velocity field can be modeled
as fully-developed, stationary turbulence.  Gravitational and thermal
instabilities under the influence of galactic rotation 
contribute to form the turbulent velocity field. 
The Toomre effective 
$Q$ value, in the non-linear phase, ranges over a wide
range, and gravitationally stable and unstable regions are distributed
patchily in the disk.
These results suggest that large-scale
galactic rotation coupled with the self-gravity of the gas
can be the ultimate energy sources that maintain the turbulence in the
local ISM.
We find that our models of turbulent rotating disks are
consistent with the velocity dispersion of an extended HI disk in the
dwarf galaxy, NGC 2915, where there is no prominent
active star formation.  Numerical simulations show that the stellar
bar in NGC 2915 enhances the velocity dispersion, and it also drives
spiral arms as observed in the HI disk.
\end{abstract}


%%%%%%%%%%%%%%%%%%%%%%%%%%%%%%%%%%%%%%%%%%%%%%%%%%%%%%%%%%%%%%%%%%%%%
%
\section{INTRODUCTION}
%
%%%%%%%%%%%%%%%%%%%%%%%%%%%%%%%%%%%%%%%%%%%%%%%%%%%%%%%%%%%%%%%%%%%%%
Observational and theoretical
studies have suggested that there are supersonic hydrodynamical or
magneto-hydrodynamical turbulent motions in molecular clouds and also
in various phases of the ISM \citep{mccray79,larson81,quiro83,balbus91,gold95,gold97,franco99}.
  A number of numerical studies
showed, however, that the turbulence in the clouds decays as
$t^{-\eta}$, with $ \eta \sim 1$ \citep{mac98, ostriker01}.
The dissipation time of the turbulence is of the order of the flow
crossing time or smaller, even in the presence of strong magnetic
fields \citep{stone98}.  These numerical experiments suggest that
energy input is necessary to maintain the turbulent motion in the
molecular clouds.  Some energy sources originating in stellar
activity have been proposed: stellar winds from young stars
\citep{norman80}, photoionization \citep{mckee89}, and supernova
explosions \citep{MO}.  However, these processes can not be the main
energy sources for keeping the turbulent motion in molecular clouds
that do not harbor stellar activities \citep{will98}.  In this case,
external shocks caused by distant stellar energy release may produce
vortices in the clouds \citep{korn00}.

In differentially rotating disks with a magnetic field, 
the magneto-rotational instability \citep{balbus91}
is a probable cause for the turbulent motion. 
\citet{sell99} have studied the MHD turbulence in extended HI disks,
applied their model to the extended HI disk of 
NGC 1058, and have showed that the observed uniform velocity dispersion 
may be modeled by the MHD-driven turbulence.

In terms of the origin of the non-self-gravitating, 
pure hydrodynamical turbulence, on the other hand,
Balbus, Hawley and Stone (1996) have 
claimed, on the basis of theoretical and numerical studies, that 
Keplerian disks are linearly and non-linearly stable. 
They rejected self-sustained hydrodynamical turbulence with
outward angular momentum transport, because
the turbulence cannot gain energy from the differential rotation.
However, \citet{kato97} suggested a steady hydrodynamical turbulence
in Keplerian disks, which are caused by the pressure-strain tensor.
Richard \& Zahn (1999) have re-analiezed laboratory experiments of 
the Couette-Taylor flow, and concluded that turbulence may be sustained by
differential rotation when $d\Omega/dR < 0$.
\cite{godon99} showed that
purely hydrodynamical perturbations can develop initially into either sheared disturbances or coherent vortices (see also Papaloizou \& Pringle 1985).
However, the perturbations decay and do not evolve into a self-sustained turbulence.

Another important mechanism that generates turbulence in 
a rotating disk
is self-gravitational, local or global
 instability of the gas \citep{gold65a,gold65b}.
Although there are some simulations of self-gravitating, 
turbulent gas in a local shearing box (e.g. V$\acute{\rm{a}}$zquez-Semadeni et al. 1995; Gammie 2001), 
%turbulent gas in a local shearing box \citep{vaz95,gammie01} 
the local approximation would not be adequate to study 
the nature of self-gravity dominated turbulence \citep{balbus99}. 
The global hydrodynamical simulations given by  \citet{LKA2} have 
revealed spiral unstable modes in the differentially rotating disk,
but the development of self-sustained turbulence was not observed.

In the this paper, we study the gravity-driven turbulence in 
galactic disks using two-dimensional, global hydrodynamical
simulations.  We solve
the basic hydrodynamical equations and the Poisson equation
numerically, taking into account realistic radiative cooling.  We use
the same numerical technique discussed by Wada \& Norman (1999,2001) (hereafter
WN99, WN01), in which kpc-scale dynamics of 
the multi-phase ISM can be followed with a sub-pc resolution.
WN99 and WN01 suggested that a gravitationally unstable and thermally 
unstable disk can evolve into a globally quasi-stable disk with
quasi-stationary turbulent velocity field. Here we perform 
simulations with a higher spatial resolution, and make
a detailed analysis of the turbulent structure.

 In \S 2, we briefly summarize the numerical method and models, and
in \S 3, we show that turbulent energy spectra are achieved over a
wide dynamic range (from kpc to pc) in the direct numerical
experiments.  In order to study the effects of pure turbulence in the
disk, we here ignore the energy feedback due to supernovae
and stellar winds from massive stars \citep{NF96}.
In \S 4, we discuss what is required for 
realistic models of the ISM, and we apply our model to 
the HI disk of the dwarf galaxy, NGC 2915. 
Extended HI disks are particularly
relevant objects to compare with our models, because they are not
significantly affected by the star formation \citep{dickey90,meurer96}. 
We summarize our results in \S 5.

%%%%%%%%%%%%%%%%%%%%%%%%%%%%%%%%%%%%%%%%%%%%%%%%%%%%%%%%%%%%%%%%%%%%%
%
%\section{NUMERICAL SIMULATIONS OF THE TURBULENT INTERSTELLAR MEDIAM}
\section{NUMERICAL METHODS AND MODELS}
%
%%%%%%%%%%%%%%%%%%%%%%%%%%%%%%%%%%%%%%%%%%%%%%%%%%%%%%%%%%%%%%%%%%%%%

The numerical methods are the same as those described in WN99 and
WN01.  Here we summarize them briefly.  We solved the following
equations numerically in two-dimensions:
\begin{eqnarray}
\frac{\partial \Sigma_g}{\partial t} + \nabla \cdot (\Sigma_g \mbf{v}) &=& 0, \label{eqn: rho} \label{eq1} \\ 
\frac{\partial \mbf{v}}{\partial t} + (\mbf{v} \cdot \nabla)\mbf{v}
+\frac{\nabla p}{\Sigma_g} &=& -\nabla \Phi_{\rm ext} - \nabla \Phi_{\rm sg}  , \label{eqn: rhov}  \label{eq2} \\  
\frac{\partial E}{\partial t} + \frac{1}{\Sigma_g} \nabla \cdot [(\Sigma_g E+p)\mbf{v}] &=& \Gamma_{\rm UV}- \Sigma_g \Lambda(T_{\rm g}), \label{eq3}\\
\nabla^2 \Phi_{\rm sg} &=& 4 \pi G \Sigma_g \delta(z) \label{eqn: poi}
\end{eqnarray}
where, $\Sigma_g,p,\mbf{v}$ are surface density, pressure, and velocity of the gas, 
the specific total energy $E \equiv |\mbf{v}|^2/2+ p H/[(\gamma -1)\Sigma_g]$
with $\gamma = 5/3$ and a scale height $H$.
The potential has two parameters, $a$ (scale parameter called the `core radius') and $v_{\rm max}$ (the maximum rotation speed), and is expressed as
\begin{equation}
\Phi_0 (R)=  \frac{c^2}{a {(R^2 + a^2)}^{1/2} } \  ,  \label{eq-1}
\end{equation}
 where $c $ is given as  $c \equiv v_{\rm max}
{(27/4) }^{1/4} a $, with $v_{\rm max} = 150$ km s$^{-1}$ and $a=$0.2 and
2 kpc. 

We also assume a cooling function $\Lambda(T_g) $ $(10 < T_g < 10^8
{\rm K})$ with Solar metallicity and a heating due to
photoelectric heating, $\Gamma_{\rm UV}$. A constant scale height ($H = 100$ pc)
is assumed to calculate the cooling rate.
We assume a uniform UV radiation field,
equal to the the local Galactic UV field \citep{ger97}. 
However, the UV heating does not significantly affect the dynamics of the
ISM. 

The hydrodynamic part of the basic equations is solved by AUSM
 (Advection Upstream Splitting Method), \citep{LS}. 
We achieve third-order spatial accuracy with MUSCL \citep{VL}. 
The limiter function is chosen to satisfy 
the TVD (Total Variation Diminishing) condition. 
In this scheme, an explicit numerical viscous term is not required
and shocks are captured sharply.
We used $4096^2$ equally spaced Cartesian grid points covering a 2 kpc
$\times $ 2 kpc region around the galactic center. 
We also run models using $1024^2$ and $2048^2$ grid points to
evaluate how numerical resolution affects the results.

The Poisson equation is solved to
calculate the self-gravity of the gas using the FFT and the
convolution method \citep{HE}.  The second-order leap-frog method is used for the
time integration.  We adopt implicit time integration for the cooling
term.

  The initial condition is an axisymmetric and rotationally
supported disk assuming the Toomre'$Q$ parameter, $Q = 1.2$.
%The total mass of the gas is ?? $M_\odot$, which is ? \% of the dynamical mass.
Random density and temperature fluctuations are added at each grid cell
in the initial disk. These
fluctuations are less than 1 \% of the unperturbed values and have an
approximately white noise (flat power spectrum) distribution.
%%%%%%%%%%%%%%%%%%%%%%%%%%%%%%%%%%%%%%%%%%%%%%%%%%%%%%%%%%%%%%%%%%%%%
%
\section{RESULTS}
%
%%%%%%%%%%%%%%%%%%%%%%%%%%%%%%%%%%%%%%%%%%%%%%%%%%%%%%%%%%%%%%%%%%%%%
As described in WN01, thermal and gravitational instabilities grow
non-linearly over the whole of the disk in a few dynamical time
scales.  The whole system shows a quasi-stable state after $t \sim 20
$ Myr (2 rotational periods at $R = 0.2$ kpc).  As shown in Fig. 1,
which is a density field of a part of the disk (500 pc $\times$ 500 pc, i.e. 
1/16 of the entire simulated region),  a complicated network
of clumps and filaments as well as low density voids are formed.  The
density ranges from $10^{-4}$ to $10^{6}$ \mpc.  The velocity field in
the same region in Fig. 1 is also very complicated as shown in Fig. 2a and 
2b, in which  $\nabla \cdot \mbf{v}$ and $(\nabla \times \mbf{v})_z$ 
are plotted.  Comparing to Fig. 1 and 2, we see that regions of
converging velocity roughly coincide with dense filaments, and
positive and negative vortices are also associated with the filaments.
This shows that local shearing motions are generated along the filaments.
One should note that this turbulent velocity field is maintained in a
rotating, self-gravitating gas disk without any explicit energy
inputs, such as supernovae or any other artificial driving forces.

In order to understand the nature of the turbulent velocity field
statistically, we calculate the energy spectra, $E(k)$ from the
velocity field, where $k$ is the wave number.
  We decompose the velocity field into two components,
i.e. $\mbf{v} = \mbf{v}_{\rm comp} + \mbf{v}_{\rm sol}$, where
$\mbf{v}_{\rm comp}$ is the compressible velocity field and
$\mbf{v}_{\rm sol}$ is the incompressible (or solenoidal) field
\citep{passot88}.  The two components are defined from $\nabla \cdot
\mbf{v}_{\rm sol} = 0$ and $\nabla \times\mbf{v}_{\rm comp} = 0$, and
solenoidal and compressible energy spectra, $E_s(k)$ and $E_c(k)$ are
calculated from $\mbf{v}_{\rm sol}$ and $\mbf{v}_{\rm comp}$,
respectively.  Figure 3a shows the evolution of the compressible
energy spectrum, $E_c(k)$. 
 One finds that the compressible part of the
velocity field 
reaches a quasi-steady state in a few rotational periods, where the
spectrum shows a double power-law shape with a `knee' at $k\sim 300$
 ($\lambda \sim 7$ pc).
The kinetic energy is mainly driven at wavenumbers around $k_i \sim
200-300$. The turbulence inversely cascades to a larger scale ($k
\lesssim 100$), and simultaneously it also cascades to smaller scales.
The input wave number, $k_i$, corresponds to about 10 pc, which is the
scale of the initial gravitational instability in the cool phase after
the initial cooling.  The inverse cascade implies a hierarchical growth
of the gravitational instability in the disk.  The energy spectrum of
the solenoidal component $E_s(k)$ similarly evolves as that $E_c(k)$
does (Fig. 3b), but it has more power especially on the large scale
($k \lesssim 10$ or $\lambda \gtrsim 200$ pc)
 because of the galactic rotation.  The appearance of
the `hump' at $k\sim 200$ at $t=1$ Myr and the following evolution
suggest that eddies are generated on the same scale as the
gravitational instability, and they propagate toward large and small
scales. The rotational
energy on a larger scale cascades downward at $k \sim 10$.
This process is just like evolution of the gravitational
instability in a rotating disk \citep{gold65a,gold65b}.
We discuss this below.

A stationary power-law regime in the energy spectrum of the system
implies a fully-developed turbulence, and its power-law index
represents some statistical aspects of the turbulence.  Before
discussing the power-law indices and the range of the inertial
cascade, we first check the simulation for numerical artifacts. Since the turbulence of the ISM is a phenomenon over a wide dynamic range, one should be 
careful about fitting the numerically obtained 
spectrum with a power-law.  We are
using a non-adaptive Eulerian grid, therefore 
the results on smaller scales can 
be affected by the numerical resolution.  We therefore check the
dependence of the turbulent spectrum on the resolution by using
$2048^2$ and $1024^2$ grid points as well as the $4096^2$ grid points
(i.e. 0.49 pc resolution).  In Fig. 4, $E(k) = E_s(k) +
E_c(k)$ for three cases using $4096^2, 2048^2,$ and $1024^2$ grid
points at $t=20.2$ are plotted.  We find that 1) the three runs coincide at $k
\lesssim 100$ within a factor of two, 2) the slopes of the `inertial
ranges' are the same $d {\rm ln} E(k)/d {\rm ln} k \sim -0.8$, and
3) the power-law indices of the `dissipation regions' in the all
three models approach to $\sim -5.2$.  The positions of the `knees' are
shifted by about factor of two, therefore the inertial range 
is increased by a factor of two if we use twice  the
number of grid points.  Therefore, we can safely conclude that the
slope of the inertial range is not strongly affected by the
spatial resolution.
%We should be careful that 
%their absolute values depend on the
%resolution, especially for $k \gtrsim 100$. With higher resolution,
%the local velocity fields gain more energy}.

In fully developed Kolmogorov incompressible turbulence, the steady
power-law energy spectrum indicates a steady energy flow from large
scales toward smaller scales.  In two-dimensional, incompressible,
hydrodynamical turbulence, 
stationary spectra are found to be proportional to $k^{-3}$
for $k > k_f$ (enstrophy, (i.e. squared vorticity) cascade regime) and
to $k^{-5/3}$ for $k < k_f$ (energy `inverse' cascade regime), where
$k_f$ is the forcing wavenumber \citep{kra67}. This inverse energy
cascade is also observed in numerical simulations of a weakly
compressible 2-D turbulence \citep{dah90}.   The evolution of
the spectrum in Fig. 3a look consistent with the
Kraichnan's picture, with $k_f \sim 200$, although 
our spectrum is shallower ($d {\rm ln} E_s(k)/d{\rm ln}k \sim -1$) in 
the inertial range than the Kraichnan's prediction. 

In Fig. 5, the final energy spectra ($t=48$ Myr), $E(k)$, $E_s(k)$,
 and $E_c(k)$ for the model using $4096^2$ grid points are plotted.
 $E_c(k)$ has a power-law part $E_c(k) \propto k^{-0.8}$ between
 $k\sim 20$ and 200, and $E_s(k) \propto k^{-1}$ between $k\sim 10$ and 200.
 It is known that for shock-dominated, fully-developed turbulence,
 $E(k) \propto k^{-2}$, and for a weakly shocked turbulence, $E_c(k)
 \propto k^{-1.5}$ are expected \citep{passot88}. The present spectrum
 is shallower than these values, and it implies that the structure of
 velocity field is not determined only by shocks.  However, it is not
 clear whether this difference originates from the complexity of the
 present system, i.e.  a non-uniform, rotating, self-gravitating, and
 radiative turbulence.  
Even so, it is interesting that the resultant spectra in
this complicated system show stationary power laws over inertial ranges as
seen in the much simpler systems - uniform, non-selfgravitating, and
non-radiative turbulence.
Note that the slope will change if the
 simulations are generalized to include energy feedback from
 supernovae, which cause strong shocks in the ISM (WN01).

Next, we show energy spectra for two models with different rotation
curves, i.e. rigid and differential rotation curves ($a = 2$ and
$a=0.2$ kpc) in Fig. 6.
The plot shows that the spectra of the rigid rotation model on large scales ($k < 10$) are smaller than those of the differential rotation model by a factor $10-100$. On the other hand, the difference of the spectra between the two models are a factor $2-3$ on small scales ($k > 10$). 
Interestingly, the spectra of the inertial range in the rigid rotation model are steeper ($d{\rm ln}E(k)/d{\rm ln k} \sim -1.5$).
It is reasonable that 
energy extracted from the galactic rotation 
is more effective when there is a shear field.
We discuss this more quantitatively below.

The above numerical results suggest that the ISM in galaxies is
dynamic, but in a statistical sense, it is in a `quasi-steady state',
whose velocity field resembles fully developed turbulence over a
wide dynamic range (at least three orders of magnitude).  The expected
velocity dispersions are $\sim 10$ km s$^{-1}$ and 0.1 km s$^{-1}$ for
100 pc and 1 pc scales, respectively.  There are two obvious energy
sources that can maintain the turbulence in this system: the
shear-driven by the galactic rotation and local tidal field due to the
self-gravity of the gas.  Besides these, local pressure gradients can
also contribute to generate the turbulence.  This is because that the
turbulent ISM is no longer in pressure equilibrium (WN01, Gazol et
al. 2001).  As shown by the energy spectra (Fig. 5), the rotational
energy is dominated by that on a large scale ($ > 200$ pc).  On the
other hand, turbulence is first driven by the self-gravity of the gas on a
scale of 10--20 pc as seen in Fig. 3.  These energy inputs are in
equilibrium with the energy transport due to turbulent decay and 
losses due to the radiative cooling. 

In order to see the effect of the gravitational instability on the 
generation of the turbulence, we calculate the {\it effective} 
Toomre stability
parameter $Q \equiv \kappa \sigma_e/\pi G \Sigma_g$, where
the effective velocity dispersion, $\sigma_e$, is defined by
 $\sigma_e^2 \equiv \sigma^2 + c_s^2$
with the velocity dispersion $\sigma$ and the sound velocity $c_s$,
the average surface gas density $\Sigma_g$ and, the epicyclic frequency,
$\kappa$. All quantities are averaged in local subregions (1 kpc$/\Delta R = 
100$ and $2\pi/\Delta \phi = 60$).
 Fig. 7 shows the volume-weighted 
distribution functions of the $Q(R,\phi)$ for
three different snapshots.
It is notable that $Q$ is distributed over a wide range, i.e. four-orders of 
magnitudes. At $t=12.6$ Myr, a large fraction of the volume is
in the unstable ($Q < 1$) state. Note that,  although the $Q < 1 $
implies that
the axisymmetric mode is linearly unstable, the disk can be
unstable to non-axisymmetric
modes even with $Q > 1$. 
The peaks shift to larger values for later times, which
means the disk is stabilized. At $t=44.2$ Myr, the median is $Q \sim 5$,
and the regions where $Q \ll 1$ and $Q \gg 1$ coexist. 
This situation is clearly seen in Fig. 8, which shows the 2-D distribution
of $Q(R,\phi)$ and $\Sigma_g(R,\phi)$ at $t=44.2$ Myr.
The distribution of the stable (yellow-white) and unstable regions
(dark red-blue-black) are very patchy. It is obvious that 
the unstable regions correspond
to high density regions, which are seen in the density map colored 
red-yellow. 
The disk is more unstable in the outer region ($R \gtrsim 0.5$ kpc) than
in the inner region.
The density map shows some global spiral structures, 
which can be also seen in the $Q$ map. The inhomogeneous 
distribution of $Q$ appears in the initial a few rotations, and then
they evolve into larger structure. This evolution represents
the development of the inhomogeneous structure in density, and 
the evolution of $E(k)$ as shown in Fig. 3 is consistent with this
result. The most unstable wave length $\lambda_m$ in the 
differentially rotating disk is $\lambda_m \equiv 2\pi^2 G \Sigma_g/\kappa^2 
\sim $3--30 pc. This is consistent with the typical size of the unstable
regions seen in Fig. 8, which is less than about 50 pc.
The sizes of the unstable regions are larger than those of the 
'dissipational region' seen in the energy spectra (Fig. 5), therefore
the patchy structure is not strongly affected by the numerical resolution.
The ISM in the $Q < 1$ regions are expected to be collapsed,
 but this does not mean that all gas components in each unstable region
eventually turn into {\it one} high density clump. In fact, 
there are substructures in the unstable regions, therefore
due to the local angular momentum, the pressure gradient, and the 
Coriolis force, the collapse cannot be monotonic.
As a result,  the gravitational contraction
 causes the local velocity dispersion and shock heating, then
the regions are stabilized. On the other hand,
the stable regions can become unstable due to mass inflow and 
radiative cooling. Therefore the local structure of the
stable and unstable regions are time-dependent, but the
entire patchy structure seen in Fig. 8 does not change, 
which is consistent with fact that the energy spectra do not evolve
significantly once the system reach the quasi-steady state.

The above results imply that the origin of the turbulent motion in 
the multi-phase, differentially rotating disk
is gravitational instability.
%The instability causes the velocity dispersion, and it stabilizes
%the global structure of the disk.
What we observe here is just like the `local' gravitational instabilities in
a rotating disk as discussed by Goldreich \& Lynden-Bell (1965a,b),
although the gas disks here have much more complicated multi-phase structures.
The turbulence is driven by the gravitational instabilities, and 
probably it is also supplied by  energy extracted from the galactic
rotation. This is 
the same mechanism that drives the MHD turbulence. As discussed by
\citet{sell99}, the energy supply rate is given by
$-T_{R\phi} d\Omega/d\ln R$, where the stress tensor is
expressed using Alfven velocity $\mbf{u}_A$ and gravitational 
velocity $\mbf{u}_G$ as
$T_{R\phi} = \langle   \rho(u_R u_\phi - u_{AR} u_{A\phi}+ u_{GR} u_{G\phi})
\rangle$ (see also Lynden-Bell \& Kalnajis 1972).
Therefore, even if there is no magnetic field, the gravitational stress
tensor works to extract energy from the galactic rotation.

The energy supply rate due to the gravitational instability
 per unit mass can be 
estimated as  $ {\dot \varepsilon} \sim \langle u_{GR} u_{G\phi} \rangle \Omega \sim G (\Sigma_g/H) \lambda^2 \Omega
\sim 5 \Omega \, ({\rm km \, s}^{-1})^2 (\Sigma_g/10 M_\odot {\rm pc}^{-2}) (\lambda/100 {\rm pc})^2 (H/100 \, {\rm pc})^{-1}$
 (where $\lambda$ is a scale length of the turbulence)
 and $H$ is the scale height of the disk.
The time scale of the energy supply, 
i.e. $v_{\rm rot}^2/{\dot \varepsilon} \sim (150 \, {\rm km} \,  {\rm s}^{-1})^2/5\Omega \sim 
700 \tau_{\rm rot}$, where $\tau_{\rm rot}$ is
the rotational period,  is long enough compared to the galactic 
life time. 
In other words, the turbulence can be sustained using a small fraction of
the galactic potential energy.
%The dissipation time scale is 	
%$3/2 k T/\rho^2 \Lambda(T) \sim ?? Myr$, and it is comparable to
%that of the energy subtraction from the galactic rotation.
The dependence of the energy supply rate on the wavelength 
can be derived roughly as 
${\dot \varepsilon} \sim G \Sigma_g/H \lambda^2 \cdot Rd\Omega/dR
\propto \lambda^{1/2}$ for the Kepler rotation ($\Omega \propto R^{-3/2}$)
or $ {\dot \varepsilon} \propto \lambda^2 \Omega$ for the rigid rotation. 
Since the turbulent energy dissipation rate per unit mass should have 
the same dependence on the turbulent scale as the energy supply rate does, 
$v_{\rm turb}^2/(\lambda/v_{\rm turb}) = v_{\rm turb}^3/\lambda \propto \lambda^{1/2}$, or 
$\propto \lambda^2$.
As a result, we expect that $E(k) \propto k^{-1}$ for
the Kepler rotation,  or  $E(k) \propto k^{-2}$ for 
the rigid rotation.
That is, the power spectrum is steeper in 
the rigid rotation case than in the differentially rotating case as we see in
Fig. 6. One should note that the turbulence extracts energy from
the galactic rotation through self-gravity, and the galactic global shear 
is not a necessary condition to sustain the turbulence.

In addition to the local instability in self-gravitating disks, 
interactions with the non-axisymmetric disturbance,
 such as  spiral density waves,  can be a
mechanism to redistribute energy from the background shear to
the local disturbance through the $\langle u_R u_\phi \rangle$ term in the stress tensor \citep{lynden72, papa91, LKA2}.
This has an analogy with the swing amplification of 
spiral density waves in differentially rotating, stellar disks \citep{julian66,sel84}. As seen in Fig. 8, 
global spiral density waves evolve in the inhomogeneous disks
after a few rotational times.

%%%%%%%%%%%%%%%%%%%%%%%%%%%%%%%%%%%
%
\section{DISCUSSION}

\subsection{Towards more realistic modeling of the ISM}
Modeling the ISM on a galactic scale is a
challenging problem in numerical astrophysics.  
It is required that realistic simulations include
many elementary processes, such as various radiative cooling/heating
processes, thermal conduction, interaction between the magnetic field and the ISM, 
self-gravity of the gas, energy feedback from supernovae, etc.
Moreover, the simulations should be ultimately three-dimensional and global, 
that is the whole galactic disk should be solved.
In this sense, our model and any other past numerical simulations of the ISM
are still highly idealized (see a review by V$\acute{\rm{a}}$zquez-Semadeni 2002). Our model is global, and 
realistic radiative cooling and self-gravity of the gas are
taken into account, but
the magnetic field, which is important for the evolution of the
interstellar turbulence (see references in \S 1), is ignored.
Therefore we should be careful to apply the present results to the
real ISM.
On the other hand, most MHD simulations of the ISM 
adopt the local shearing box approximation.
There are many local 2-D models, e.g. self-gravitating ISM with radiative cooling (Vazquez-Semadeni et al. 1996), and  self-gravitating ISM 
with isothermal or adiabatic equation of state \citep{kim01,gammie01}.
Some models in 3-D consider
radiative cooling and supernova feedback \citep{korpi99}, 
but most of them are local, non-selfgravitating, and an isothermal equation of 
state (EOS) is assumed \citep{mac99}. 
There was some global, 3-D MHD simulations with polytropic EOS calculated for 
the galactic central region, but they ignored self-gravity of the gas
(e.g. Machida, Hayashi, \& Matsumoto 2000). 
Using self-gravitating, global hydrodynamical simulations with radiative cooling,
on the other hand, Wada (2001) showed that the ISM in the galactic central 
region has the complicated filamentary, multi-phase structure as seen in 
the 2-D simulations presented here.
The results in the present paper and the past attempts are complementary,
and they should be improved  with more consistent numerical modeling of the ISM
in the future. 

Finally we would like to comment on
 plausibility of the 2-D approximation for the ISM.
The real ISM in a galactic
disk should behave as three-dimensional turbulence below the scale
height of the cold gas ($\sim 100$ pc).
Since the turbulence is driven by local gravitational instability, 
we expect that the ISM in 3-D behaves like 2-D
turbulence on larger scales,
therefore the inertial range of $d{\rm ln} E(k)/d{\rm ln} k \sim -1$ 
would also appear for $k \lesssim 40$ in 3-D ($k \lesssim 200$ in 2-D).
For $k \gtrsim 40$, the energy spectrum would be Kolmogorov-like, i.e.
$E(k) \propto k^{-5/3}$, or if it is shock-dominated, $E(k) \propto k^{-2}$
would be expected. In a real ISM on a smaller scale, 
the main energy sources would be
magneto-hydrodynamical instability and/or
energy feedback from stars as well as the energy decay from larger scales. 
Transition from 2-D to 3-D turbulence and its effect on the
energy spectra are now important problems that can be addressed in
studies of the gas dynamics in galactic disks, 
This would be clarified utilizing
three-dimensional, global simulations of the ISM 
with magnetic field on a galactic scale.

One might wonder if the spatial resolution in our simulation is
meaninglessly fine (i.e. $\sim 0.5$ pc), if we apply the model to
the real ISM whose scale height is $\sim 100$ pc.
However, as shown in Fig. 4, 
the transition from the inertia range to the `dissipative' part
is affected by the numerical resolution. The maximum wave length
of the inertia range is about 10 times larger than 
the grid size. Therefore even for 2-D simulations, the
spatial resolution should be much finer than 
the assumed scale heigh of the disk.

\subsection{Origin of Velocity Dispersion in the HI Disk of NGC 2915}
%
%%%%%%%%%%%%%%%%%%%%%%%%%%%%%%%%%%%
The HI disk in the dwarf galaxy, NGC 2915, extends to
over five times the Holmberg radius, and there is no active star
formation observed outside the central region. Therefore, this galaxy
is relevant to our study of the dynamics of the turbulent gas disk
without the influence of the stellar energy feedback.
Meurer, Mackie, \& Carignan (1994) and Meurer et al. (1996) observed
NGC 2915 ($D = 5.3$ Mpc, $M_B = -15.9$ mag, $R_{\rm Ho}=2.93$ kpc)
using the ATCA (Australia Telescope Compact Array) with a linear
resolution is 640 pc  and 
the AAT (Anglo-Astrallia Telescope).  The total mass of the HI disk is $9.6\times
10^8 M_\sun$ and $R_{\rm HI} = 14.9$ kpc. The integrated
star-formation rate is 0.05 $M_\sun$ yr$^{-1}$, and most of the
star-formation is near the center of the galaxy.  The HI
intensity map shows that there are spiral arms extending well beyond
the optical extent.  The isovelocity contours show that the disk is
rotating with a small amount of non-circular motion.  The
line-of-sight velocity dispersion of the HI is $\sim 20-40$ km
s$^{-1}$ inside the optical radius, and it is $\sim 8$ km s$^{-1}$ in
the extended HI disk. The central question is what causes this large
velocity dispersion in the extended HI disk? 

We apply the same numerical method in \S 2 to model the HI disk in NGC
2915 to obtain the radial distribution of the velocity dispersion in a
non-star forming disk.
We scale the potential model (eq.(5)) and the gas disk
to represent the HI disk of NGC 2915. Two rotation curves are assumed
for an axisymmetric component of the external potential: 1) the
potential derived from the HI rotation curve (Model A in Fig. 16 of
Meurer et al. 1996), where the core radius $a$ in eq.(5) is 4 kpc, and
2) models with a smaller core radius ($a = 2$ kpc).  The latter model
was chosen, because the HI rotation curves derived from tracing peak
intensities of position-velocity maps do not necessarily represent
{\it true} mass distribution of the galaxies, especially for the
central part \citep{sofue99}.  Sofue et al. (1999) claimed that the
true central rotation curves tend to be steeper than those implied
from the peak intensities in PV maps.

NGC 2915 has a central optical bar.  The
observed spiral patterns could be resonance-driven structures by this
central bar.  
The velocity dispersion in the disk is
generally enhanced by the non-axisymmetric potential and its
resonances.  
The pattern speed of the bar can be directly measured \citep{tre84}, and it is $\Omega_p = 8.0\pm 2.4$
km s$^{-1}$ kpc$^{-1}$ \citep{bureau99}.
In order to investigate the effects of the central bar, we also
performed runs with a non-axisymmetric potential (bar-potential) with
the same pattern speed as observed (Bureau et al. 1999), and where the
length of the bar is taken to be equal to the core radius $a$.  The
non-axisymmetric part of the potential is assumed to be in the form $
\Phi_1 (R, \phi, t) = \varepsilon (R) \Phi_0 \cos 2(\phi -\Omega_{\rm
p} t) \ , $ where $\varepsilon (R) $ is given as $ \varepsilon(R)=
\varepsilon _0 {aR^2}/{(R^2+a^2)^{3/2}}.  $ The parameter
$\varepsilon_0$ represents the strength of the bar \citep{wad01c}.
The gas is initially distributed in an axisymmetric, 15 kpc radius
disk with a exponential-like surface density profile to resemble the
observed HI distribution (Fig. 15 in Meurer et al.  1996).
We have run models with parameters 
as $6 \le \Omega_p \le 11 $ km s$^{-1}$ kpc$^{-1}$,
$0.05 \le \epsilon_0 \le 0.15$, and $a = 2$ and 4 kpc.

 The velocity dispersions of an axisymmetric model $A$ ($a=4$ kpc)
at $t=5.4$ Gyr are plotted as a function of the radius in Fig. 9a.
The velocity field is sampled at every 20 grid points (i.e. 292 pc)
for $x$ or $y$-directions, and they are averaged in the same size as
the observed beam size to calculate the dispersion.
We find that the dispersion is about a factor 4 smaller than the
observed value (20-40 km s$^{-1}$) at $R < 2$ kpc, a factor 2 smaller
than the observed 8 km s$^{-1}$ at $R > 2$ kpc.  The velocity
dispersions of a bar model $B$ ($a= 2$ kpc, $\varepsilon_0 = 0.1,
\Omega_p = 8$ km s$^{-1}$ kpc$^{-1}$) is plotted in Fig. 9b.  The
central velocity dispersion in this model is $\sim 20-30$ km s$^{-1}$,
and the it is about 3--4 km s$^{-1}$ in the outer disk.  The central
velocity dispersion is comparable with the observed value, but the
dispersion in the outer disk is still factor 2 smaller than
observations.

Fig. 10 shows $E(k)$ for the
two models $A$ and $B$. Both spectra show double power-law shapes,
$E(k) \propto k^{-1}$ and $k^{-1.5}$ in $ 10 \lesssim k \lesssim 100$
for models $A$ and $B$, respectively. 
The slope of the axisymmetric model is comparable to that in 
Fig. 5. The steeper slope in the bar model would be due to stronger shocks
caused by the bar. The spectra also shows that 
the bar model has a factor of 2-10 larger energy than the
axisymmetric model in the most scales. 

In Fig. 11 and 12, we plot the density and PV maps of the bar model
 $B$.  Two major spiral arms, which are formed from many spirals and
 clumps, are evident as well as an inhomogeneous compact disk.  The PV
 diagram shows a large velocity dispersion at $R < 3$ kpc, which
 corresponds to the clumpy nuclear disk.

The Toomre's $Q$ parameter estimated by the observations 
is 5--9 \citep{meurer96}. 
However, this does {\it not}  mean that there are no local gravitationally 
unstable regions in the HI disk.  
The beam size in the observations is much larger than the typical
size of the unstable regions. 
Therefore, we can observe the global stability of the HI disk, but
probably the local instability is observed as the velocity dispersion 
in the beam size.

After exploring the parameter space ($\epsilon_0,\, \Omega_p$, and $a$),
we conclude that a set of parameters, i.e. 
$\epsilon_0 \sim 0.1, \Omega_p \sim 8 $ km s$^{-1}$
kpc$^{-1}$, and $a\sim 2$ kpc, reasonably reproduces the observed radial
distribution of the velocity dispersion, and the HI morphology.
However, the observed value of the velocity dispersion in the extended
disk is about factor two larger than that in the best-fit model $B$.
This might be due to the limitation of our model (\S 4.1),
or because of the following reasons.
Fig. 10 suggests that stronger bars can enhance 
the velocity dispersion in the disk by about factor of two. If the
observed velocity dispersion is a result of {\it hidden} 
gravitational instability in the HI disk,
the dispersion could be larger in a more massive disk. That
is, the observed HI mass does not necessarily represent the total mass
of the gas in the galaxy.  There might be a component of high density,
compact molecular clouds in the extended HI disk, which could be
observed by the next-generation radio interferometer, ALMA.

\section{SUMMARY}
%
%%%%%%%%%%%%%%%%%%%%%%%%%%%%%%%%%%%
Dynamics and structure of the interstellar medium (ISM) in galactic
disks are numerically studied, using high-resolution, 2-D
hydrodynamical simulations with a large dynamic range.  In \S 3, we
showed that the velocity fields of the ISM in a steady-state show
turbulent-like spectral energy distributions which are natural
consequences of the non-linear development of 
gravitational instabilities of the ISM.  The energy spectra are
obtained over three-orders of magnitude in the wave number.
The model using
$4096^2$ grid points shows that the spectrum for the compressible and
incompressible (solenoidal) part of the velocity field in the
inertia ranges are fitted approximately
as $E(k) \propto k^{-1.5}$ and $k^{-1}$, respectively.  We found that
the turbulent energy spectra are achieved in a few dynamical
time-scales, and they subsequently remain in a quasi-steady state. We
do not include energy feedback from supernovae or other external
driving forces to produce and maintain turbulence. The turbulence is
self-maintained over the wide dynamic range. From analyzing the
temporal evolution of the energy spectra and the density structure, we
suspect that energy input from self-gravitational interaction of the
ISM and the galactic rotation drives the turbulence balancing the
turbulent energy decay and the radiative cooling.

In \S 4, we compared our global ISM models with HI observations of 
NGC 2915. A large velocity dispersion ($\sim 8$ km
s$^{-1}$) is observed in the extended HI disk of NGC 2915, even though
there is no active star formation.  We found that the observed
velocity dispersion and its radial distribution in NGC 2915 can be
quantitatively understood as the gravity-driven turbulence, 
but the galactic potential
requires an additional non-axisymmetric potential with a rotation
speed of $\Omega_p \sim 8$ km s$^{-1}$ kpc$^{-1}$ or 
the total gas mass should be larger than the observed HI mass.

%%%%%%%%%%%%%%%%%%%%%%%%%%%%%%%%%%%
%
\acknowledgments 
%
%%%%%%%%%%%%%%%%%%%%%%%%%%%%%%%%%%%
The authors thank Fabian Heitsch for his valuable comments on 
the draft. 
Numerical computations were carried out on Fujitsu VPP5000 at NAOJ.  
KW is supported in part by Grant-in-Aids for Scientific Research 
(no. 12740128) of JSPS.

\clearpage
%fig1
\begin{figure}
      \epsscale{0.5}
\plotone{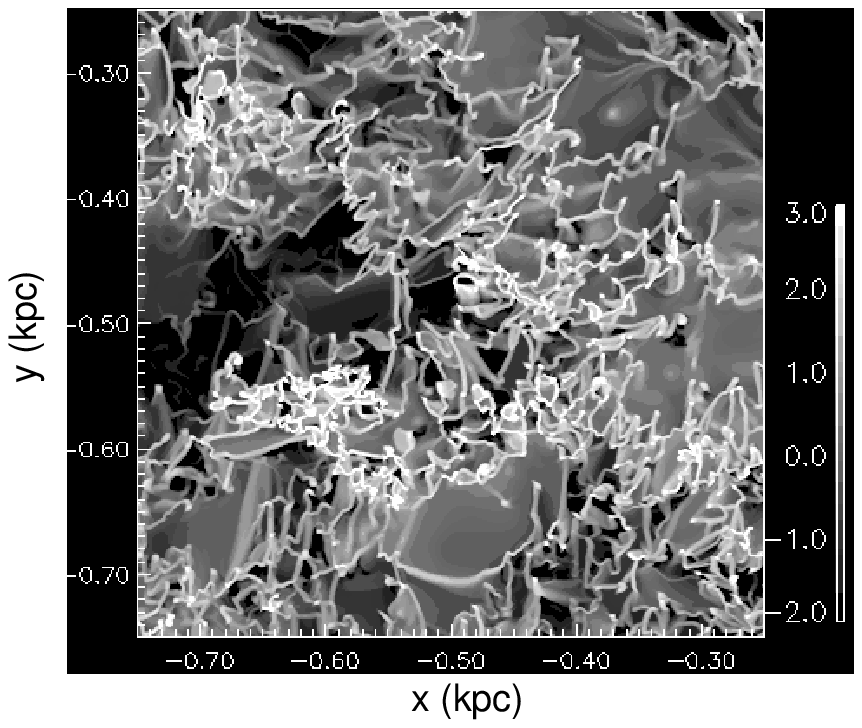}
%      \plottwo{epsfile}{epsfile}
\caption{Density distribution of the gas in $1/16$ 
of the whole calculating region (2 kpc $\times$ 2 kpc) at $t=48.2$.
$(x,y) = (0,0)$ is the galactic center. The gray scale represents
log-scaled density ($M_\odot $ pc$^{-2}$).}
\end{figure}

%fig2
\begin{figure}
      \epsscale{0.9}
\plotone{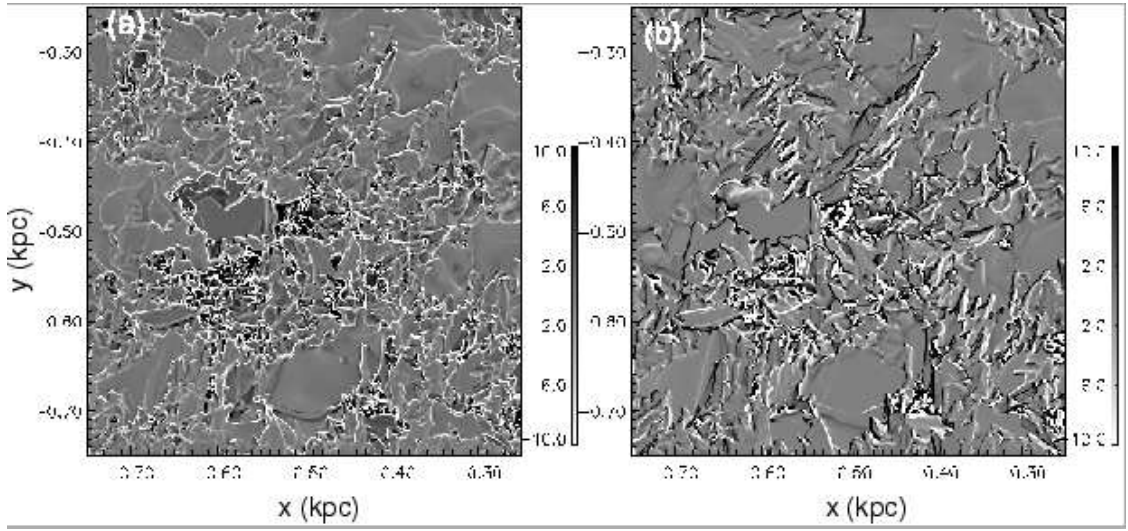}
	\caption{(a) Divergence of the velocity field ($\nabla \cdot \mbf{v}$) 
in the same area of Fig. 1. (b) Same as (a), but for $z$-component of 
rotation, i.e. $(\nabla \times \mbf{v})_z$.
 The unit is km s$^{-1}$ pc$^{-1}$. }
\end{figure}

%fig3
\begin{figure}
      \epsscale{0.9}
\plotone{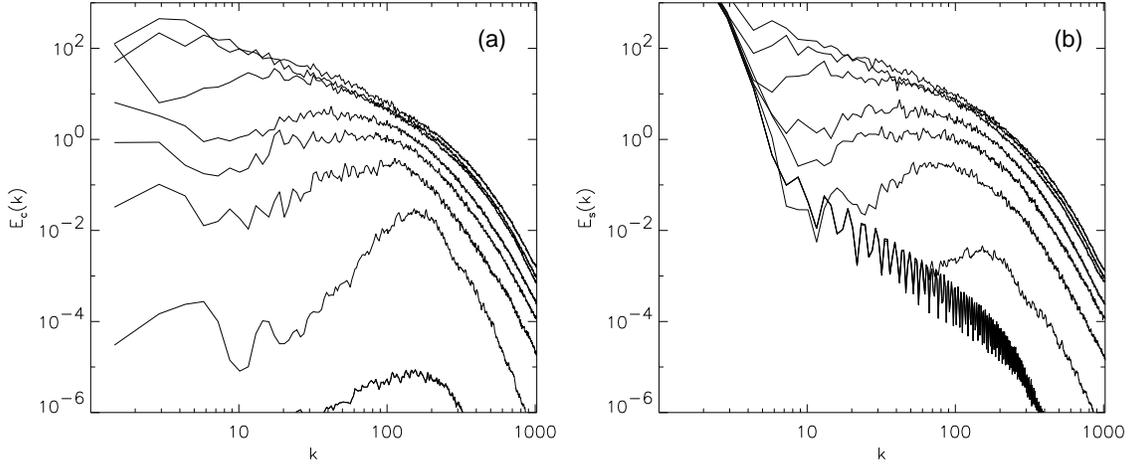}
\caption{ (a) Evolution of energy spectrum for the compressible
part of the velocity field, $E_c(k)$ is 
plotted for  $t = 0.2, 1, 2, 3, 4, 8, 16$, and 36 Myr.
(b) Same as (a), but for the solenoidal part, $E_s(k)$.}
\end{figure}

%fig4
\begin{figure}
      \epsscale{0.5}
\plotone{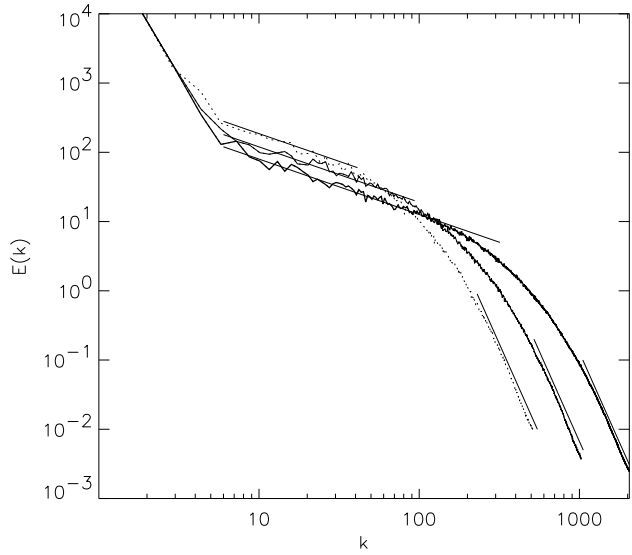}
\caption{Energy spectra, $E(k) = E_s(k) + E_c(k)$,  for the three same 
models at $t=20.2$, but with different spatial resolutions: 
$4096^2$ (thick solid curve), $2048^2$ (thin solid curve), and $1024^2$ grid points are used (dotted curve). The six thin solid lines 
represent power-law whose slopes are -0.8 and -5.2.}
\end{figure}

%fig5
\begin{figure}
      \epsscale{0.5}
\plotone{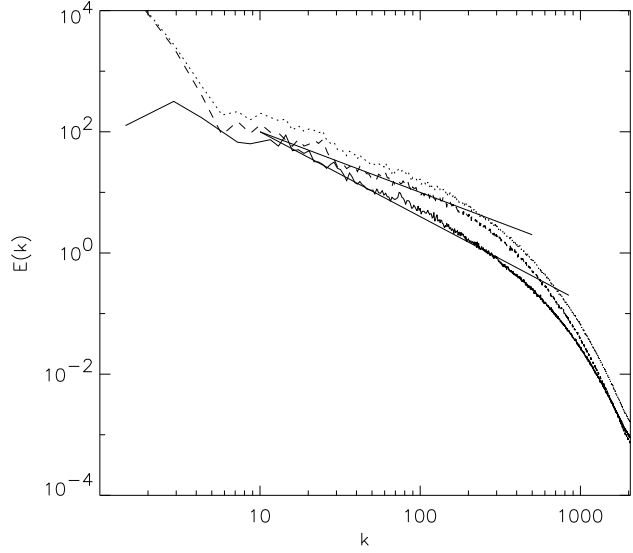}

\caption{Energy spectra, $E(k)$ (dotted line), $E_s(k)$ (dashed line),
 and $E_c(k)$ (solid line) of 
the high resolution model ($4096^2$ grid points are used) at $t=$ 48 Myr.
The two thin solid lines represent power-laws of
 $k^{-1.5}$ and $k^{-0.8}$.}
\end{figure}

%fig6
\begin{figure}
      \epsscale{0.5}
\plotone{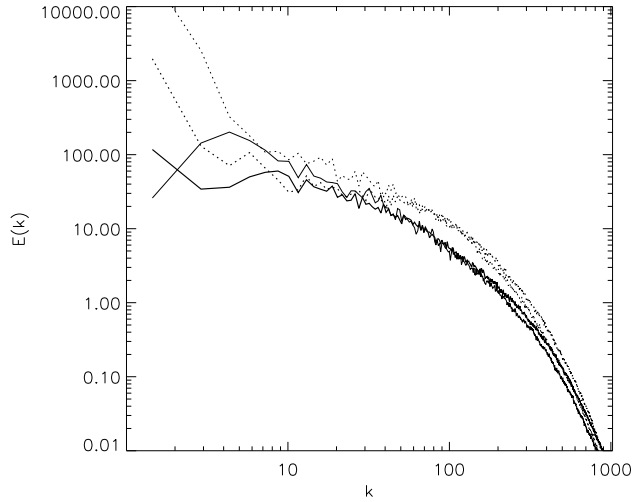}

\caption{Energy spectra in two models with 
different rotation curves. The thick and thin
solid lines are $E_c(k)$ and $E_s(k)$ for the ridge rotation (the core radius, $a=2$ kpc). 
The two dotted lines are the same, but for the differential rotation ($a=0.2$ kpc). 
The number of grid points are $2048^2$.}
\end{figure}

%fig7
\begin{figure}
      \epsscale{0.5}
\plotone{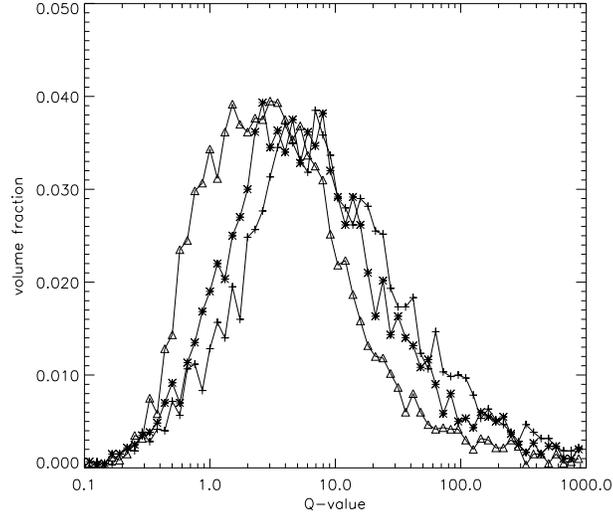}

\caption{Histogram of the 'effective' Toomre's $Q$ parameter 
at $t=$ 12.6 (triangles), 25.6 (asterisks), and 44.2 Myr (crosses)
of the high resolution model ($4096^2$ zones).}
\end{figure}

%fig8
\begin{figure}
      \epsscale{0.5}
\plotone{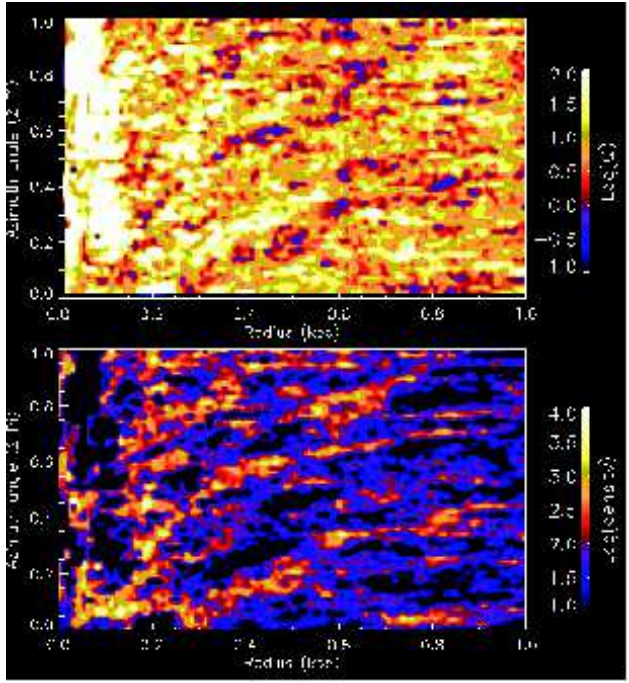}
\caption{Two-dimensional distribution of the effective $Q (r,\phi)$ 
(upper panel) and the density $\Sigma_g (r, \phi)$ (lower panel)
at $t=48$ Myr of
the high resolution model ($4096^2$ zones). All quantities to calculate
the effective $Q$, i.e. the epicyclic frequency, 
surface density and velocity dispersion are averaged in 100 $\times$ 60 sub-regions.}
\end{figure}

%fig9
\begin{figure}
      \epsscale{0.9}
\plotone{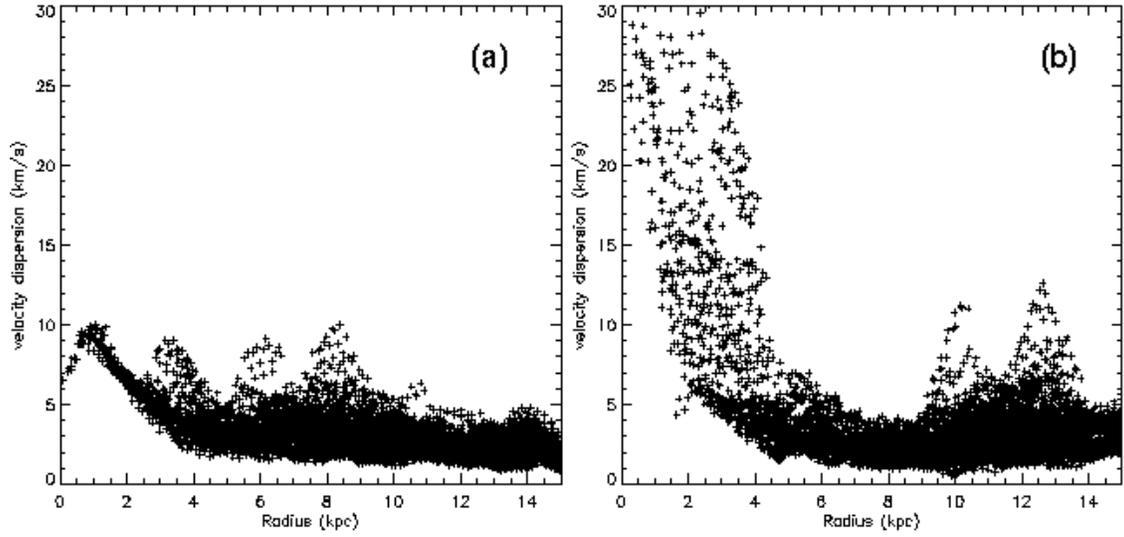}
\caption{Radial distribution of the velocity dispersion for
(a) axisymmetric model $A$, and (b) bar model $B$ for the extended HI disk of 
NGC 2915.}
\end{figure}

%fig10
\begin{figure}
      \epsscale{0.5}
\plotone{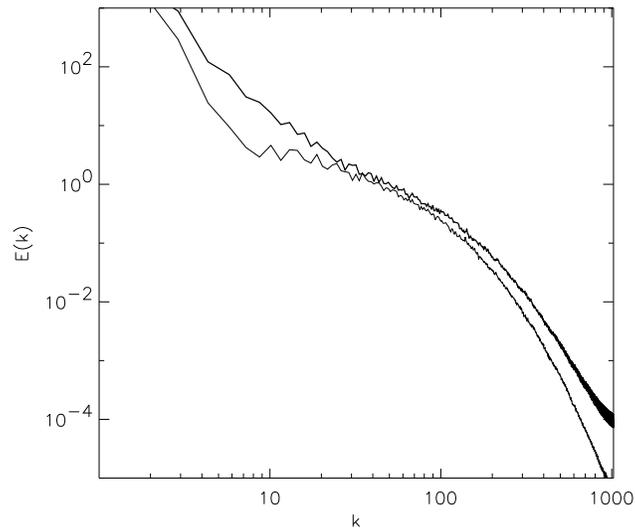}

\caption{Energy spectra for the axisymmetric model $A$ (thin line) and 
for the bar model $B$ (thick line).}
\end{figure}

%fig11
\begin{figure}
      \epsscale{0.5}
\plotone{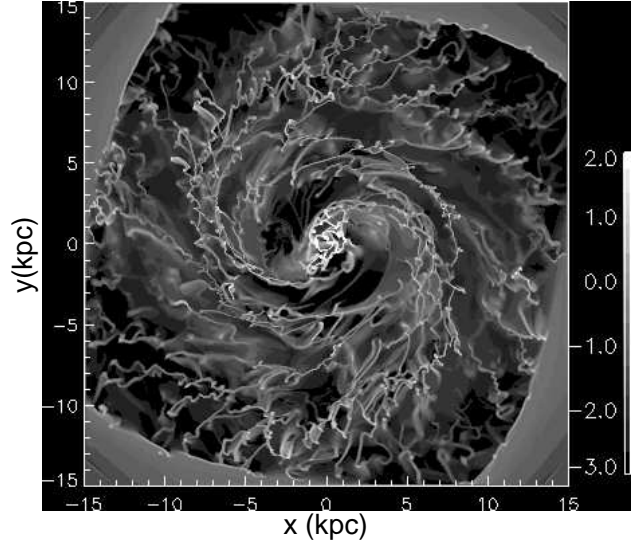}

\caption{Density distribution of model $B$. 
The two major spirals are formed due to the outer Lindblad resonance,
which is caused by the central stellar bar of the
pattern speed 8 km s$^{-1}$ kpc$^{-1}$.
}
\end{figure}

%fig12
\begin{figure}
      \epsscale{0.7}
\plotone{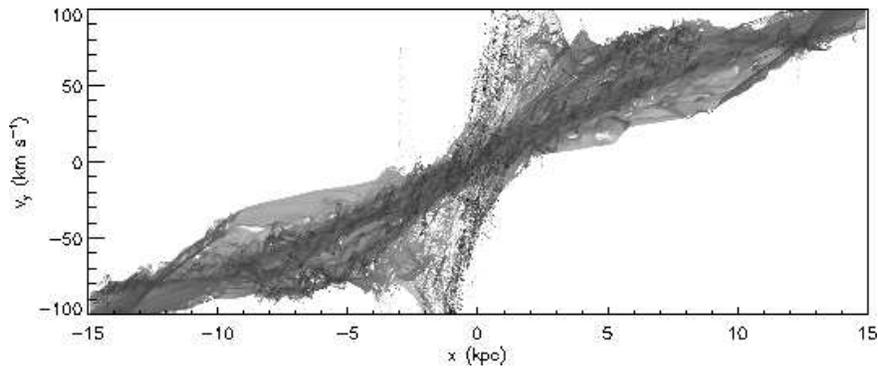}

\figcaption{Position-Velocity ($x-v_y$) diagram of model $B$.
The central high velocity components originate in the central
clumpy disk seen in Fig. 11.}
\end{figure}

\end{document}